\documentclass[twocolumn,showpacs,preprintnumbers,aps,pra,amsmath,amssymb]{revtex4}
\usepackage{graphicx}
\usepackage{dcolumn}
\usepackage{bm}
\usepackage{amssymb}
\usepackage[german, english]{babel}
\begin{document}
\draft

\title{Collective oscillations of a Fermi gas in the unitarity limit:\\Temperature effects and the role of pair correlations}

\author{S. Riedl,$^{1,2}$
E. R. {S\'{a}nchez Guajardo},$^{1}$ C. Kohstall,$^{1}$ A.
Altmeyer,$^{1,2}$ M. J. Wright,$^{1}$ J. {Hecker Denschlag},$^{1}$
and R. Grimm$^{1,2}$}

\address{$^{1}$Institut f\"ur Experimentalphysik und Zentrum f\"ur Quantenphysik,
Universit\"at Innsbruck, 6020 Innsbruck, Austria\\
$^{2}$Institut f\"ur Quantenoptik und Quanteninformation,
\"Osterreichische Akademie der Wissenschaften, 6020 Innsbruck,
Austria}

\author {G.\ M.\  Bruun$^{3,4}$ and H.\ Smith$^4$}
\address{$^3$Dipartimento di
Fisica, Universit\`a di Trento and CNR-INFM BEC Center, I-38050
Povo, Trento, Italy\\
$^4$Niels Bohr Institute, University of Copenhagen,
DK-2100 Copenhagen \O, Denmark}

\date{\today}


\begin{abstract}
We present detailed measurements of the frequency and damping of
three different collective modes in an ultracold trapped Fermi gas
of $^6$Li atoms with resonantly tuned interactions. The measurements
are carried out over a wide range of temperatures. We focus on the
unitarity limit, where the scattering length is much greater than
all other relevant length scales. The results are compared to
theoretical calculations that take into account Pauli blocking and
pair correlations in the normal state above the critical temperature
for superfluidity. We show that these two effects nearly compensate each
other and the behavior of the gas is close to the one of a classical
gas.

\end{abstract}
\maketitle
\section{Introduction\label{collmod}}
The study of collective oscillations in quantum liquids and gases
has yielded a wealth of insights into the properties of strongly
correlated systems. An early example concerning strongly correlated
Fermions is the observed transition from ordinary first sound to
zero sound in the normal state of liquid $^3$He as the temperature
is lowered~\cite{Abel1966poz}. In this Article we explore related
phenomena in an ultracold quantum gas of fermions in the unitarity
limit~\cite{Inguscio2006ufg} by measuring three different collective
modes under similar conditions. The frequency and damping of the
modes exhibit the characteristic transition from hydrodynamic
behavior at low temperature to collisionless behavior at higher
temperature. The experimental observations are compared to
theoretical model calculations that apply to the normal state of the
gas above the critical temperature $T_c$ for superfluidity. In the
unitarity limit, the strongly correlated normal state between $T_c$
and the Fermi temperature $T_F$ is arguably not as well understood
as the $T=0$ superfluid phase~\cite{Giorgini2007tou}. It is shown
that the combined effects of temperature and pair correlations
account for most of the observed features in this interesting
temperature regime.

\begin{figure}
\includegraphics[width=7cm]{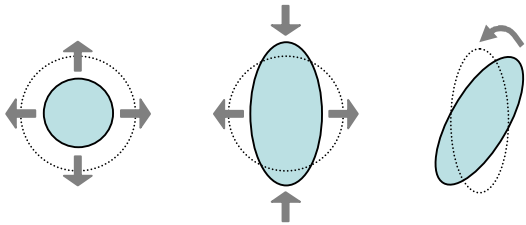}
\caption{\label{dance} Sketch of the three  collective modes
investigated in this work: the compression mode, the quadrupole
mode and the  scissors mode (from left to right). The oscillations
take place in the plane of tight confinement, perpendicular to the
direction of the elongated, cigar-shaped cloud. While the compression mode represents an
oscillation of the overall cloud volume, the other two modes only
involve surface deformations. Exciting the quadrupole mode leads
to an oscillating elliptic shape. The scissors mode appears as an
angular oscillation of an elliptic cloud about a principal axis of
an elliptic trap geometry.} \label{fig1}
\end{figure}

Our measurements of the collective modes are carried out for an
elongated trap geometry, which has previously been shown to be well
suited for studying  the dynamical behavior of a strongly
interacting Fermi gas
\cite{Bartenstein2004ceo,Kinast2004efs,Kinast2004boh,Kinast2005doa,Altmeyer2007pmo,Altmeyer2007doa,Wright2007ftc}.
We focus on two collective excitations of a cylindrically symmetric
cigar-shaped cloud, namely the radial compression mode and the
radial quadrupole mode. In addition we study the scissors mode under
conditions where the cloud exhibits pronounced  ellipticity in the
plane perpendicular to the direction of the cigar-shaped cloud. In
all three modes, the cloud oscillates mainly in the plane normal to
the direction of the cigar-shaped cloud. For a sketch of the modes
see Figure \ref{dance}.

Previous experiments on collective modes in a strongly interacting
Fermi gas studied the effect of the interaction strength in the zero
temperature limit.
\cite{Bartenstein2004ceo,Kinast2004efs,Kinast2004boh,Altmeyer2007pmo,Altmeyer2007doa}.
Systematic investigations were performed studying the radial
compression mode
\cite{Bartenstein2004ceo,Kinast2004efs,Altmeyer2007pmo} and the
radial quadrupole mode \cite{Altmeyer2007doa}. Measurements on the
compression mode served as a sensitive probe for the equation of
state of the gas in the zero temperature limit throughout the
BEC-BCS crossover regime. In contrast to the compression mode, the
frequency of the radial quadrupole mode allows one to test the
hydrodynamic behavior without being influenced by the equation of
state. This made it possible to investigate the transition from hydrodynamic
to collisionless behavior with decreasing coupling strength of the
atom pairs on the BCS side of the crossover.

While the hydrodynamic behavior in the zero-temperature limit is now
well understood as a result of superfluidity, an understanding of
the effects of temperature on the collective modes has remained a
challenge. Only few experiments have so far addressed this problem
\cite{Altmeyer2007pmo,Kinast2005doa,Wright2007ftc,Kinast2004efs}.
Previously, the temperature dependence of the radial compression
mode \cite{Kinast2005doa} and the scissors mode was studied
\cite{Wright2007ftc}. Our present experiments aim at addressing the
open questions raised by the different results obtained in these
experiments: The frequency and damping of the radial compression
mode was studied as function of the temperature in an experiment
performed at Duke University \cite{Kinast2005doa}. There the mode
frequency appeared to stay close to the hydrodynamic value even for
temperatures exceeding the Fermi temperature. This surprising
finding stands in contrast to scissors mode measurements, performed
later at Innsbruck University \cite{Wright2007ftc}, which clearly
showed a transition to collisionless behavior in the same
temperature range. Furthermore the Duke data on the damping of the
compression mode did not show a maximum as it was seen in the
Innsbruck data on the scissors mode measurement. These apparent
discrepancies are a particular motivation for our present study of
different collective modes under similar experimental conditions.

\section{Experimental Procedure\label{experiment}}\label{collmode}
The apparatus and the basic preparation methods for experiments with
a strongly interacting Fermi gas of $^6$Li atoms have been described
in our previous work \cite{Jochim2003bec,Bartenstein2004cfa}. As a
starting point, we produce a molecular BEC of $^6$Li$_2$. By
changing an external magnetic field, we can control the
interparticle interactions in the vicinity of a Feshbach resonance,
which is centered at 834\,G \cite{Bartenstein2005pdo}. The
measurements of the collective modes are performed at the center of
the Feshbach resonance, where the interactions are unitarity
limited.

The atoms are confined in an elongated, nearly harmonic trapping
potential, where the trap frequencies $\omega_x$ and $\omega_y$ in
the transverse direction are much larger than the axial trap
frequency $\omega_z$. The confinement in the transverse direction is
created by an optical dipole trap using a focused 1030\,nm laser
beam with a waist of 47\,$\mu$m. Note that the Gaussian shape of the
laser beam leads to significant anharmonicities in the trapping
potential. The potential in the axial direction consists of a
combination of optical and magnetic confinement; the magnetic
confinement is dominant under the conditions of the present
experiments. The trap parameters, given in Table
\ref{modeparameters}, represent a compromise between trap stability
and anharmonic effects \cite{parameters}. The Fermi temperature is
given by $T_F=E_F/k$, where the Fermi energy
$E_F=\hbar(3N\omega_x\omega_y\omega_z )^{1/3}=\hbar^2 k_F^2/2m$,
$k_F$ is the Fermi wavenumber and $k$ is the Boltzmann constant. The
parameter $V_0$ is the trap depth, and $N$ is the total number of
atoms, given by  $N=6\times10^5$. The interactions are characterized
by the dimensionless parameter $1/k_Fa$, where $a$ is the $s$-wave
scattering length.

\begin{table}
\caption{\label{modeparameters}Trap parameters for the different
modes.}
\begin{ruledtabular}
\begin{tabular}{cccc}
 &compression&quadrupole&scissors\\
\hline
$\omega_x/2\pi$ (Hz) & 1100 & 1800 &1600\\
$\omega_y/2\pi$ (Hz) & 1100 & 1800 & 700\\
$\omega_z/2\pi$ (Hz) & 26 & 32 & 30\\
$T_F$ ($\mu$K) & 1.8 & 2.7 & 1.9\\
$V_0/k$ ($\mu$K) & 19 & 50 & 40\\
\end{tabular}
\end{ruledtabular}
\end{table}

To control the aspect ratio $\omega_x/\omega_y$, we use rapid
spatial modulation of the trapping beam by two acousto-optical
deflectors, resulting in the creation of time-averaged trapping
potentials \cite{Altmeyer2007doa}. This on one hand allows us to
compensate for residual ellipticity of the trapping potential on the
percent level and thus to realize a cylindrical symmetric trap
$(\omega_x = \omega_y)$. On the other hand it allows for the
excitation of surface modes by deliberately introducing elliptic
trapping potentials $(\omega_x \neq \omega_y)$. The procedures used
to excite the modes are outlined in Appendix A. To change the
temperature we apply a controlled heating scheme via sudden
compression of the gas as described in \cite{Wright2007ftc}.
Detection of the cloud is done by absorption imaging which displays
the shape of the cloud in the $x$-$y$ plane after expansion. For
each mode under investigation we determine the frequency and damping
following the procedures of our previous work
\cite{Altmeyer2007doa,Wright2007ftc,Altmeyer2007pmo}; see also
Appendix A.

\begin{figure}
    \includegraphics[width=7cm]{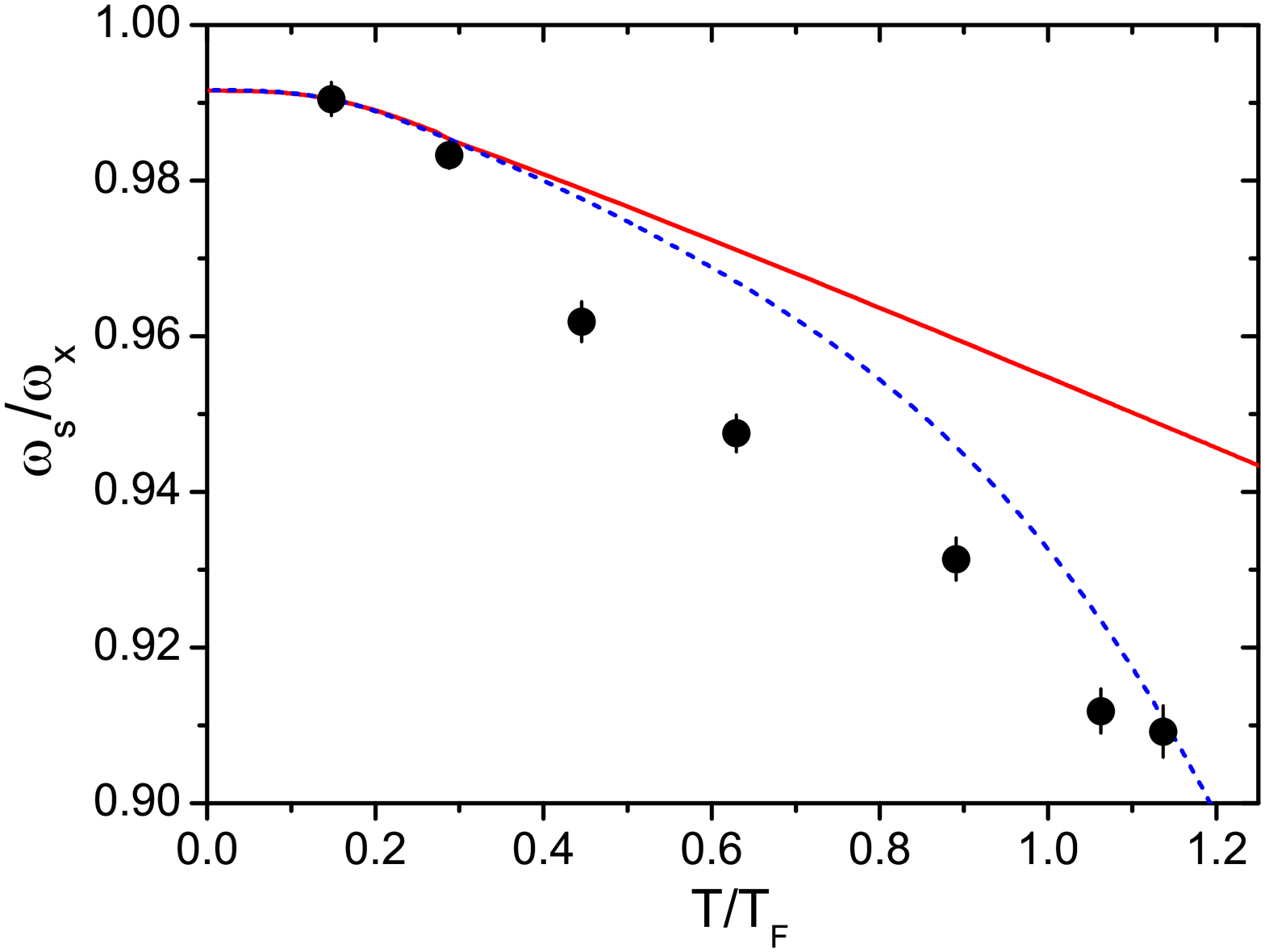}\\
  \caption{Sloshing mode frequency $\omega_s$ normalized by the trap
  frequency $\omega_x$
  as a function of temperature. The measured frequency shows a
  decrease with increasing temperature,
  (dots) which is due to the increase in the size
  of the cloud. The lines show the expected frequency from a
   first-order anharmonic correction; see Appendix B. To determine
   the cloud size for different temperatures we assume a harmonic potential (solid line) and a Gaussian
   potential (dashed line), respectively \cite{profiles}.
  }\label{slosh}
\end{figure}

Because of the Gaussian shape of the trapping potential, corrections
are needed for a precise comparison of the experimental observation
to the idealized case of perfect harmonic trapping. Especially for
higher temperatures, when the size of the cloud is larger,
anharmonic corrections become important. This is demonstrated by
measurements of the transverse sloshing mode frequency $\omega_{s}$
(Fig.~\ref{slosh}), which clearly show a substantial decrease with
increasing temperature. To reduce the anharmonic effects on the
frequencies of the collective modes under investigation, we
normalize the compression and quadrupole collective mode frequencies
to the sloshing mode frequency in the transverse direction. This
normalization reduces the anharmonic effects to a large extent since
the decrease of the sloshing mode frequency with increasing cloud
size is of the same order as the corresponding decrease of the
frequency of the transverse modes \cite{Sringariprivat}. To
normalize the scissors mode frequency we take the geometric average
of the two different sloshing mode frequencies in the transverse
direction.

For each of the trap parameters of the different modes we determine
the sloshing mode frequency as a function of the temperature. As an
example we show $\omega_s$ for the trap parameters used for the
compression mode measurement; see Fig.~\ref{slosh}. We compare
$\omega_s/\omega_x$ (dots) to a theoretical model which allows to
calculate the sloshing frequency as a function of the cloud size;
see Appendix B. Assuming a harmonic potential to derive the mean
squared size $\langle x^2\rangle$ \cite{profiles} underestimates the
anharmonic effects (solid line) in particular for higher
temperatures. Taking into account a Gaussian potential to determine
$\langle x^2\rangle$ (dashed line) agrees much better with the
measured sloshing frequency.

Since the purpose of this article is the comparative study of
different collective modes and not the precision measurement of a
single mode as in previous work \cite{Altmeyer2007pmo}, we follow a
faster yet simpler procedure to normalize the frequencies. We
measure the sloshing mode frequency only at particular temperatures
of interest. From these points we determine the temperature
dependence of the sloshing frequency by interpolation. Even though
the normalization takes into account the temperature dependence of
the anharmonicity, it does not reduce effects due to drifts in the
power of the trapping beam. We believe this to be the main source
for the scatter of the data in Fig.~\ref{freq}.

To determine the temperature of the gas we first adiabatically
change the magnetic field to $1132$\,G \cite{technical}, where
$1/k_Fa \approx -1$, to reduce the effect of interactions on the
density distribution \cite{Luo2007mot}. Under this condition, for
$T>0.2T_F$, the interaction effect on the density distribution is
sufficiently weak to treat the gas as a non-interacting one to
determine the temperature from time-of-flight images. We fit the
density distribution after 2\,ms release from the trap to a
finite-temperature Thomas--Fermi profile. The temperature measured
at $1132$\,G is converted to the temperature in the unitarity regime
under the assumption that the conversion takes place isentropically,
following the approach of Ref. \cite{Chen2005toi}. Statistical
uncertainties for the temperature stay well below $0.05 T_F$.

\section{Theory\label{theory}}
We shall compare our experimental findings to the results of model
calculations that apply to the normal state of the gas, i.e.\ at
temperatures above $T_c$. In this Section, we outline our
theoretical approach to the calculation of mode frequencies for
$T>T_c$. A more detailed description can be found in
Refs.~\cite{Massignan2005vra} and \cite{Bruun2005vat}. We assume
that single-particle excitations are reasonably well defined in the
sense that most of the spectral weight of the single-particle
spectral function is located at a peak corresponding to that of
non-interacting particles. The low-energy dynamics of the gas can
then be described by a semiclassical distribution function
$f({\bf{r}},{\bf{p}},t)$ which satisfies the Boltzmann equation. A
collective mode corresponds to a deviation $\delta f=f-f^0$ away
from the equilibrium distribution $f^0({\mathbf{r}},{\mathbf{p}})$.
Writing
 $\delta f({\mathbf{r}},{\mathbf{p}},t)=f^0({\mathbf{r}},{\mathbf{p}})[1-f^0({\mathbf{r}},{\mathbf{p}})]\Phi({\mathbf{r}},{\mathbf{p}},t)$
and linearizing the Boltzmann equation in $\delta f({\mathbf{r}},{\mathbf{p}},t)$ yields
\begin{equation}
f^0(1-f^0)\left(\frac{\partial \Phi}{\partial t} +\dot{\bf
r}\cdot\frac{\partial \Phi}{\partial{\bf r}} +\dot{\bf
p}\cdot\frac{\partial \Phi}{\partial{\bf
p}}\right)=-I[\Phi]\label{Boltzmann},
\end{equation}
where $\dot{\bf r}={\bf v}={\bf p}/m$, $\dot{\bf p}=-\partial
V/\partial{\bf r}$ and $I$ is the collision integral. We take the
potential $V(\bf r )$ to be harmonic and given by $V({\bf
r})=m(\omega_x^2x^2+\omega_y^2y^2+\omega_z^2z^2)/2$.

To describe the collective modes we expand the deviation function
in a set of basis functions $\phi_i$ according to
\begin{equation}
\Phi({\mathbf{r}},{\mathbf{p}},t)=e^{-i\omega t}
\sum_ic_i\phi_i({\mathbf{r}},{\mathbf{p}})
 \label{Expansion},
\end{equation}
where $\omega$ is the mode frequency.  For the compression mode with
a velocity field ${\mathbf{v}}\propto(x,y,cz)$, with $c$ a constant,
we use the functions
\begin{equation}
\phi_1=x^2+y^2,\; \phi_2=xp_x+yp_y,\; \phi_3=p_x^2+p_y^2, \; \phi_4=p_z^2.
\label{breathbasis}
\end{equation}
For the quadrupole mode with a velocity field ${\bf
v}\propto(x,-y,0)$ (ignoring the small velocity along the axial
direction), we use
\begin{equation}
\phi_1=x^2-y^2,\; \phi_2=xp_x-yp_y,\; \phi_3=p_x^2-p_y^2,
\label{quadrubasis}
\end{equation}
whereas the basis functions for the scissors mode are given in
Ref.~\cite{Bruun2007fad}. Our choice of basis functions is
physically motivated by the characteristic features of the three
different modes illustrated in Fig.~\ref{dance}. Since we limit
ourselves to a few simple functions, the basis sets are not
complete, but we do not expect qualitative changes to occur as a
result of including more basis functions in our calculation.

We now insert the expansion (\ref{Expansion}) into
(\ref{Boltzmann}) and take moments by multiplying with the
functions $\phi_i$ and integrating over both ${\mathbf{r}}$ and
${\mathbf{p}}$. This yields a set of linear equations for the
coefficients $c_i$ for each of the collective modes. The
corresponding determinants give the mode frequencies. For the
compression mode, we obtain
\begin{equation}
i\omega(\omega^2-4\omega_\perp^2)+\frac{1}{\tau}\left(\frac{10}{3}\omega_\perp^2-\omega^2\right)=0
 \label{breathing},
 \end{equation}
and for the quadrupole mode, we get
\begin{equation}
i\omega(\omega^2-4\omega_\perp^2)+\frac{1}{\tau}\left(2\omega_\perp^2-\omega^2\right)=0.
 \label{quadrupole}
 \end{equation}
The equation for the scissors mode is given in Ref.~\cite{Bruun2007fad}.

The effective collision rate $1/\tau$ in (\ref{breathing}) and (\ref{quadrupole}) is given by
\begin{equation}
\frac{1}{\tau}=\frac{\int d^3{r}d^3{p}p_xp_yI[p_xp_y]}{\int
d^3{r}d^3{p}p_x^2p_y^2f^0(1-f^0)}.
\label{tau}
\end{equation}
Note that this expression for $1/\tau$  involves a spatial average
over the cloud. In the collisionless limit, $\omega\tau\gg 1$, the
two equations (\ref{breathing}) and (\ref{quadrupole}) both yield
$\omega=2\omega_\perp$, where $\omega_\perp=\omega_x=\omega_y$,
 while  in the hydrodynamic limit, $\omega\tau\ll 1$, they result in
$\omega=\sqrt{10/3}\omega_\perp$ for the compression mode and
$\omega=\sqrt{2}\omega_\perp$ for the quadrupole mode.

\begin{figure}
\includegraphics[width=\columnwidth,height=0.7\columnwidth,angle=0,clip=]{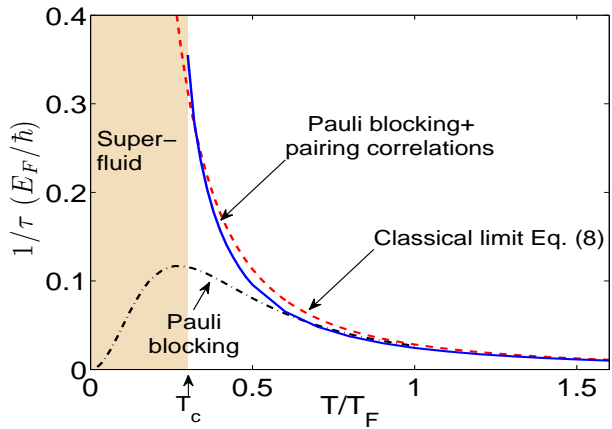}
\caption{The effective collision rate for a gas in the unitarity
limit. The dashed curve is the classical result, the dash-dotted
includes Pauli blocking, and the solid line includes pairing
correlations in the scattering matrix. The superfluid region for
$T<T_c$ is indicated.} \label{Taufig}
\end{figure}

The dependence on temperature $T$ and scattering length $a$ enters
through  $\tau$. In particular, Pauli blocking and pair correlations
strongly depend on $T$ and $a$, and we now examine their role on the
effective collision rate. In Fig.~\ref{Taufig}, we plot $1/\tau$ as
a function of temperature for a gas in the unitarity limit
$|a|\rightarrow\infty$ using three different approximations for the
collision integral.
First, the dashed curve gives the effective collision rate in the classical regime using the vacuum expression
$\mathcal{T}_{\rm vac}={\mathcal{T}}_0/(1+iqa)$ for the scattering matrix with ${\mathcal{T}}_0=4\pi\hbar^2a/m$.
The $s$-wave differential cross section  $d\sigma/d\Omega$ which enters in the collision integral $I$ is related to the
scattering $\mathcal{T}$ matrix by $d\sigma/d\Omega=m^2|{\cal{T}}|^2/(4\pi\hbar^2)^2$. In the classical regime,
we then get from (\ref{tau})
\begin{equation}
\frac{1}{\tau_{\rm class}}=\frac{4}{45\pi}\frac{kT_F}{\hbar}\frac{T_F^2}{T^2}
\label{tauclass}
\end{equation}
 for a gas in the unitarity limit~\cite{Bruun2007fad}.
 The small prefactor $4/(45\pi)\approx0.028$ in (\ref{tauclass})
 implies that the effective collision rate is significantly smaller  than what one would expect from simple estimates or dimensional
analysis at unitarity.
 Second, the dash-dotted curve gives the effective collision rate when Pauli blocking effects are
included as in~\cite{Massignan2005vra}, while
 we still use the vacuum expression $\mathcal{T}_{\rm vac}$ for the scattering matrix.
Pauli blocking effects reduce the available phase space for
scattering thereby reducing the scattering rate. For $T\ll T_F$
Pauli blocking in a normal Fermi system  gives $1/\tau\propto T^2$.
Finally, we plot as a solid curve in Fig.\ \ref{Taufig} the
effective collision rate taking into account both Pauli blocking and
many-body effects for  $\mathcal{T}$ in the ladder approximation
which includes  the Cooper (pairing) instability. This gives
${\mathcal{T}}={\mathcal{T}}_0/(1-{\mathcal{T}}_0\Pi)$ where
  $\Pi$ is the pair propagator.   Since our treatment of the pair correlations only apply to the
normal state of the gas, we plot this curve for temperatures greater than the
critical temperature  $T_c$, which within the ladder approximation
used here is given by $T_c\approx 0.3T_F$ for a
trap~\cite{Bruun2005vat}.

  We see that $1/\tau$ is  increased by the pairing correlations
 for the $\mathcal{T}$-matrix.
The pairing correlations significantly increase  the effective
collision rate for temperatures $(T-T_c)/T_c\lesssim
1$~\cite{Bruun2005vat}. One often refers to this temperature range
as the pseudogap regime.
 In fact, pairing correlations almost cancel the Pauli blocking effect in the collision integral above $T_c$ and
$1/\tau$ is fairly accurately given by the classical value as can be
seen from Fig.\ \ref{Taufig}. At high temperatures, this cancelation
can be demonstrated analytically by carrying out a high-temperature
expansion of (\ref{tau}). We obtain after some algebra the simple
expression
\begin{equation}
\frac{1}{\tau}=\frac{1}{\tau_{\rm class}}\left[1+\frac{1}{32}\left(\frac{T_F}{T}\right)^3\right].
\label{tauexpand}
\end{equation}
The presence of the small prefactor  $1/32$ in (\ref{tauexpand})
shows that the leading correction to the classical limit is less
than 3\% at temperatures above the Fermi temperature $T_F$.

\section{Results and Discussion\label{comaprison}}
\begin{figure}
\includegraphics[width=\columnwidth]{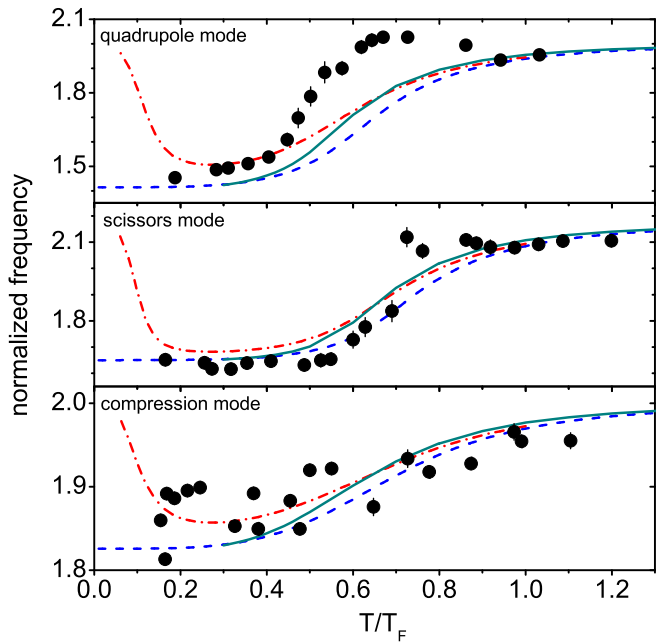}
\caption{\label{freq} The three panels show the observed normalized
mode frequencies versus temperature for the quadrupole mode, the
scissors mode and the compression mode. The error bars indicate the
statistical error of a single frequency measurement. The full lines
are the result of the theory for the normal state described in the
section \ref{theory}, which includes the combined effects
of Pauli blocking and pair correlations; note that these curves start
at $T=0.3T_F$, which in the ladder approximation used here is the
transition temperature to the superfluid state. For illustrative
purposes we also show the theoretical results when only Pauli
blocking is taken into account (dash-dotted lines) and those for a
classical gas (dashed lines). }
\end{figure}
The theoretical results of the previous section were all obtained
for a purely harmonic potential. Since anharmonicity plays an
important role in our experiments, as discussed in Sec.\
\ref{collmode}, we normalize the measured frequencies and damping
rates for the collective modes to the measured temperature-dependent
sloshing frequencies, for which an example is shown in
Fig.~\ref{slosh}. In the following we compare our observations to
the theoretical results. It should be emphasized that the
theoretical expressions for the frequency and damping contain no
free parameters to fit theory and experiment.

First we discuss the frequency for the three modes under
investigation as a function of the temperature, as plotted in
Fig.~\ref{freq}. In all three cases the theoretical expression for
the frequency (the full lines in Fig.~\ref{freq}) smoothly changes
from the hydrodynamic value at the lowest temperature considered to
the collisionless value at high temperatures.  The normalized
frequencies in the hydrodynamic limit for the quadrupole mode and
compression mode are $\sqrt{2}\approx 1.41$ and $\sqrt{10/3}\approx
1.83$, respectively. The normalized frequency in the collisionless
limit for both these modes is $2$. Using the geometric average of
the trap frequencies to normalize the scissors mode frequency, one
gets, using the ratio $\omega_x/\omega_y=16/7$ from Table I, that
$\sqrt{(\omega_x^2+\omega_y^2)/(\omega_x\omega_y)}\approx 1.65$ in
the hydrodynamic limit and
$(\omega_x+\omega_y)/\sqrt{\omega_x\omega_y}\approx 2.17$ in the
collisionless limit. Note that the scissors mode consists of a
two-frequency oscillation in the collisionless limit. Here we only
consider the larger frequency component. The lower frequency
component exhibits increasing damping towards lower temperatures and
disappears in the hydrodynamic limit \cite{Gueryodelin1999sma}.

Figure \ref{freq} illustrates that there is a reasonable  overall
agreement between experiment and theory, although some differences
exist. The agreement  is best for the scissors mode, while for the
quadrupole mode the changeover from hydrodynamic to collisionless
behavior happens at a lower temperature than the one found
theoretically. The measured compression mode frequency, which shows
considerable scatter, increases with increasing temperature and is
close to the collisionless value at the highest temperature
measured.

The observed change from the hydrodynamic to the collisionless
frequency for the compression mode is in contrast to
Ref.~\cite{Kinast2005doa}, where the frequency remains close to the
hydrodynamic value for the same temperature range. We attribute this
discrepancy to different treatments of anharmonic effects, which are
particularly important for this mode since the difference between
the hydrodynamic and collisionless frequency is of the same order as
the frequency shift due to anharmonic effects. In
Ref.\,\cite{Kinast2005doa} the data are corrected by including
anharmonic effects to first order, while we adopt the point of view
that the main anharmonic effects can be taken into account by
normalizing the measured oscillation frequencies to the measured
temperature-dependent  sloshing frequencies. Fig.~\ref{slosh}
illustrates that a simple first-order treatment of anharmonic
effects on the sloshing frequency does not account quantitatively
for the observed variation with temperature.

At very low temperatures the measured frequencies are close to the
hydrodynamic values because the gas is in the superfluid
phase~\cite{Altmeyer2007pmo,Altmeyer2007doa}. Without pair
correlations, but with Pauli blocking, at these low temperatures the
frequencies would assume their collisionless values as illustrated
by the dashed-dotted lines in Fig.~\ref{freq}.

\begin{figure}
\includegraphics[width=\columnwidth]{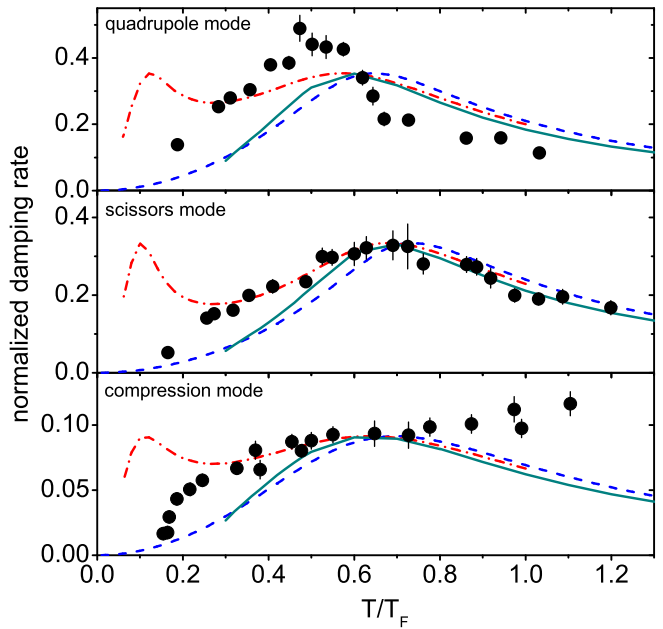}
\caption{\label{damp} Normalized mode damping versus temperature for
the quadrupole mode, the scissors mode and the compression mode. The
points are experimental values, while the full lines represent our
calculated values, taking into account both Pauli blocking and
pairing effects. The dash-dotted line only takes Pauli blocking into
account, while the dashed line is the classical (high-temperature)
result.}
\end{figure}

We now proceed to consider the damping of the oscillations. The
experimental values for the normalized damping rate are shown in
Fig.~\ref{damp}. Theoretically, one expects the damping to vanish in
the hydrodynamic and collisionless limits and exhibit a maximum in
between, as brought out by the calculations  in Sec.~\ref{theory}.
Experimentally, both the quadrupole and the scissors mode exhibit
the expected maximum in damping in the transition region. For the
compression mode, however,  the damping does not decrease at higher
temperatures. This surprising behavior for the compression mode has
already been found in \cite{Kinast2005doa}. A possible reason for
the increasing damping is dephasing-induced damping due to
anharmonicity. Anharmonic effects are  more important for the
compression mode as the intrinsic damping is relatively small due to
the small difference between the frequencies in the collisionless
and hydrodynamic limits~\cite{Gueryodelin1999coo}. In contrast to
the case of frequency discussed above, we cannot expect to take into
account the main effects of anharmonicity by normalizing the
measured damping rate to the temperature-dependent sloshing mode
frequencies. This makes it delicate to compare our experimental
results to those of a theory based on a purely harmonic potential.
The damping of the quadrupole mode shows the expected qualitative
behavior, although the maximum in damping happens at a lower
temperature compared to theory. This is consistent with  the
frequency data for this mode, since the transition there also
happens at lower temperature. For the scissors mode the experimental
data agree fairly well with theory, although some discrepancy exists
at the lowest temperatures.

We can relate the  frequency and damping of the quadrupole mode
directly to each other by eliminating the collision rate  $1/\tau$
in (\ref{quadrupole}). Writing $\omega=\omega_Q-i\Gamma_Q$ for the
solution of (\ref{quadrupole}) with $\omega_Q$ and $\Gamma_Q$ being
the quadrupole frequency and damping, we obtain
\begin{equation}
\Gamma_Q=\sqrt{-\omega_{\perp}^2-\omega_Q^2+\sqrt{8\omega_Q^2-7\omega_{\perp}^2}}.
\end{equation}

\begin{figure}
\includegraphics[width=7cm]{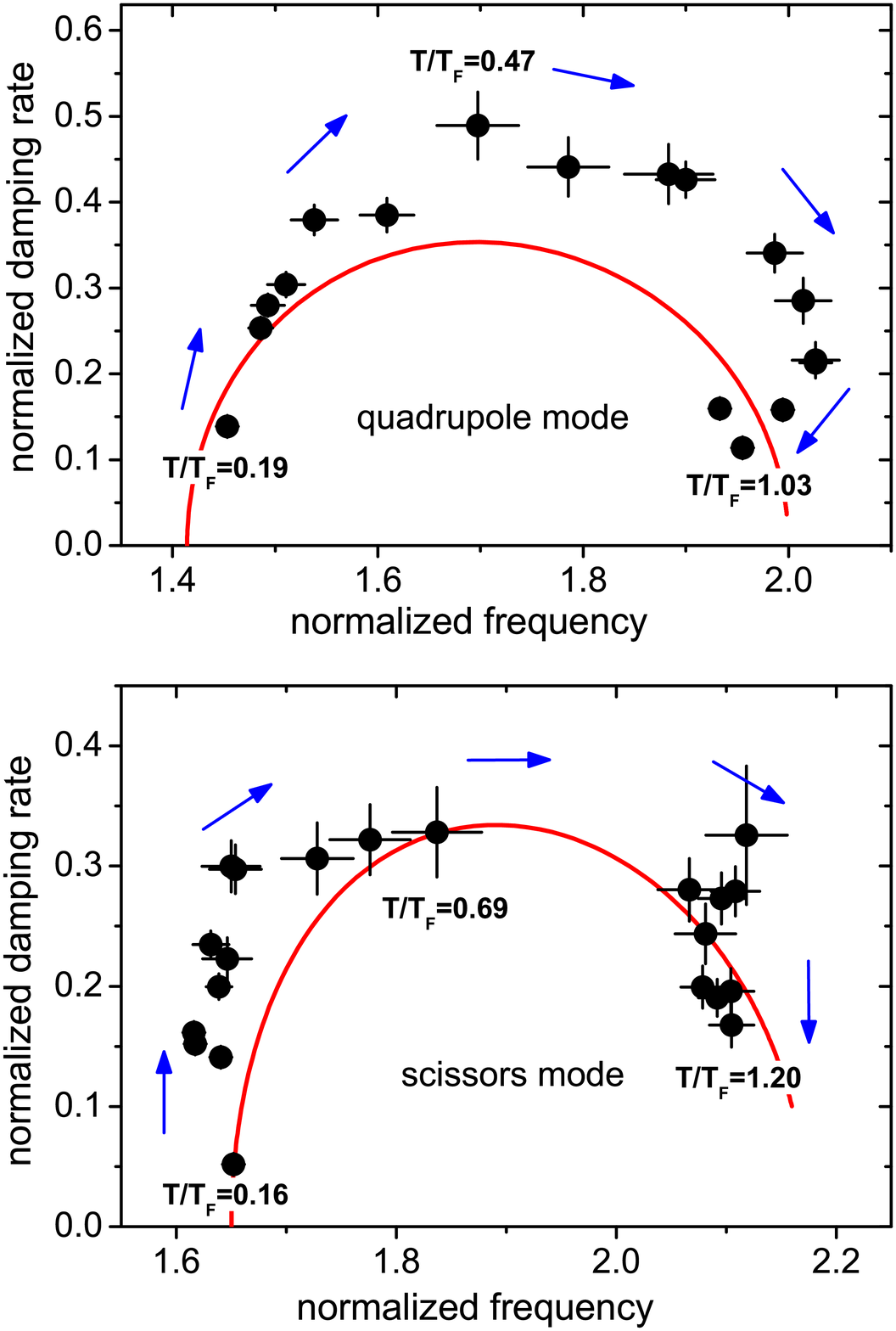}
\caption{\label{dvsf} Normalized mode damping versus normalized
frequency for the quadrupole and scissors mode. The solid line shows
the expected behavior for a harmonic trap. The arrows point toward
the direction of increasing temperature.} \label{fig6}
\end{figure}

A similar relation holds for the two other modes. This allows us to
compare theory and experiment  independently of any approximations
involved in the evaluation of  $1/\tau$. Figure \ref{dvsf} shows the
normalized damping rate versus the normalized frequency of the
quadrupole and the scissors mode; we do not show the data for the
compression mode because of the apparent problems discussed before.
We find that the maximum damping of the quadrupole mode is larger
than expected. For the scissors mode the damping is larger only at
low frequencies. This suggests that the difference between theory
and experiment is not a consequence of the approximations entering
the calculation of the relaxation rate but could be due to
anharmonic effects or the need for larger basis sets to describe the
modes [see (\ref{breathbasis}) and (\ref{quadrubasis})].

\section{Conclusion}
In this work we have presented measurements of the frequency and
damping of three different collective modes under similar conditions
for an ultracold Fermi gas  of $^6$Li atoms in the unitarity limit.
The experimental results obtained in the normal state of the gas are
in reasonable agreement with our theoretical calculations, which
take into account Pauli blocking and pair correlations. The
remaining discrepancies may originate in a variety of sources such
as our treatment of anharmonic effects, the temperature calibration,
and the use of a restricted basis for solving the Boltzmann
equation. Also they may reflect the need to incorporate further
interaction effects in the kinetic equation, which forms the
starting point for the theoretical calculations. For instance, there
are self-energy shifts on the left-hand side of the kinetic equation
which could be important. The study of collective modes is a
sensitive probe of the properties of strongly interacting particles
such as the gas of $^6$Li atoms under investigation, and further
work on temperature-dependent phenomena will undoubtedly shed more
light on these interesting many-body systems.

\begin{acknowledgments}
We acknowledge support by the Austrian Science Fund (FWF) within SFB
15 (project part 21). M.J.W.\ was supported by a Marie Curie
Incoming International Fellowship within the 6th European Community
Framework Program. Fruitful discussions with S. Stringari are
appreciated. We thank Q. Chen and K. Levin for providing us with
density profiles and temperature calibration curves.
\end{acknowledgments}

\appendix
\section{}
Here we present more details on the experimental procedures to excite the three collective modes.

To excite the radial quadrupole mode we adiabatically deform the
radially symmetric trap to an elliptic shape while keeping the
average trap frequency constant before turning off the deformation
suddenly \cite{Altmeyer2007doa}. The deformation is chosen such that
the amplitude of the mode oscillation relative to the cloud size is
below $10$\%. A two-dimensional Thomas-Fermi profile is fitted to
the images, taken after a short expansion time of $0.5$\,ms. The
difference in the width of the main axes is determined for different
hold times and fitted to a damped sine function, from which we
determine the frequency and damping of the mode.

The excitation of the radial compression mode is done by a sudden
compression of the cloud. To determine the frequency and damping of
the compression mode we follow the same procedure as for the
quadrupole mode but fitting to the sum of the widths. Here we use an
expansion time of $2$\,ms before taking the image.

The scissors mode appears as an angular oscillation of an elliptic
cloud about a principal axis of an elliptic trap. To excite this
oscillation we create an elliptic trap in the x-y plain and suddenly
rotate the angle of the principal axes by 5 degrees
\cite{Wright2007ftc}. The tilt of the principal axes of the cloud is
determined $0.8$\,ms after releasing the cloud from the trap for
different hold times. If the gas is hydrodynamic, we fit a single
damped sine function to the oscillation of the angle. However, for a
collisionless gas, the oscillation exhibits two frequencies. Thus we
fit a sum of two damped sine functions each with their own free
parameters. When the behavior changes from hydrodynamic to
collisionless the single damped sine function fits the data
reasonably well, as discussed in \cite{Wright2007ftc}. Since the
larger of the two frequencies in the collisionless regime smoothly
connects to the hydrodynamic frequency at low temperatures we only
consider this frequency in the paper.

\section{}
Here we briefly discuss the calculation of the transverse sloshing
modes including anharmonic corrections to lowest order. The
transverse trapping potential is
\begin{gather}
V(x,y)=V_0(1-e^{-x^2/a^2-y^2/b^2})\simeq\nonumber\\
V_0\left(\frac{x^2}{a^2}
+\frac{y^2}{b^2}-\frac{x^4}{2a^4}-\frac{y^4}{2b^4}-\frac{x^2y^2}{a^2b^2}
\right).
\end{gather}
Concentrating without loss of generality on the sloshing mode in the
$x$-direction, we choose the function $\Phi=c_1x+c_2p_x$. Putting
this into the linearized Boltzmann equation (\ref{Boltzmann}),
eliminating $c_2$, and taking the moment $\int dxdy n(x,y)$ with
$n(x,y)$ the density (we ignore the axial direction), we obtain for
the sloshing frequency
\begin{equation}
\omega_s^2=\omega_x^2\left(1-\frac{m\omega_x^2\langle x^2\rangle
+m\omega_y^2\langle y^2\rangle}{2V_0}\right).
\end{equation}
Here $\langle x^2\rangle=\int n(x,y)x^2dxdy/\int  n(x,y)dxdy$ and we
have used $\omega_x^2=2V_0/m a^2$ together with $\omega_y^2=2V_0/m b^2$.


\begin{thebibliography}{24}
\bibitem[{par()}]{parameters}
\bibinfo{note}{Anharmonic effects depend on the ratio between the Fermi energy
  and trap depth $E_F/V_0$ \cite{Sringariprivat}. Reducing this ratio decreases
  the anharmonic effects. This can be done by increasing the power of the
  trapping beam since $E_F$ increases more slowly than $V_0$. On the other hand
  technical reasons cause heating rates and larger drifts in the trap depth
  with increasing power.}
\bibitem[{tec()}]{technical}
\bibinfo{note}{This is the largest magnetic field for which we can take
  absorption images in our present set-up.}
\end{thebibliography}

\end{document}